# Harnessing Intrinsic Noise in Memristor Hopfield Neural Networks for Combinatorial Optimization

**Fuxi Cai[1,2,*], Suhas Kumar[1,*], Thomas Van Vaerenbergh[1,*], Rui Liu[1,3], Can Li[1], Shimeng Yu[4], Qiangfei Xia[5], J. Joshua Yang[5], Raymond Beausoleil[1], Wei Lu[2], and John Paul Strachan[1]**

[1]Hewlett Packard Labs, Palo Alto, CA, USA
[2]Department of Electrical Engineering, University of Michigan, Ann Arbor, USA
[3]Department of Electrical and Electronics Engineering, Arizona State University, Tempe, AZ, USA
[4]School of Electrical and Computer Engineering, Georgia Institute of Technology, Atlanta, GA, USA
[4]Department of Electrical and Computer Engineering, University of Massachusetts, Amherst, MA, USA
[*]Equal contributions

*We describe a hybrid analog-digital computing approach to solve important combinatorial optimization problems that leverages memristors (two-terminal nonvolatile memories). While previous memristor accelerators have had to minimize analog noise effects, we show that our optimization solver harnesses such noise as a computing resource. Here we describe a memristor- Hopfield Neural Network (mem-HNN) with massively parallel operations performed in a dense crossbar array. We provide experimental demonstrations solving NP-hard max-cut problems directly in analog crossbar arrays, and supplement this with experimentally-grounded simulations to explore scalability with problem size, providing the success probabilities, time and energy to solution, and interactions with intrinsic analog noise. Compared to fully digital approaches, and present-day quantum and optical accelerators, we forecast the mem-HNN to have over four orders of magnitude higher solution throughput per power consumption. This suggests substantially improved performance and scalability compared to current quantum annealing approaches, while operating at room temperature and taking advantage of existing CMOS technology augmented with emerging analog non-volatile memristors.*

## Introduction and background

Leveraging emerging devices such as memristors[1] to build higher performance computing systems is an attractive approach to mitigating the imminent end of Moore's law. More importantly, the challenges already faced in stalled power reduction and clock speed-up due to the end of Dennard scaling[2] focus these efforts to energy efficient systems. Very compute intensive tasks, such as the class of NP (non-deterministic polynomial)-hard problems, remain of great importance with applications in traffic management, airline scheduling, wiring within electronic chips and gene sequencing, to name a few. To address this class of problems, a variety of annealing-inspired computing accelerators using multiple technology platforms have been proposed such as adiabatic quantum annealing with super-conducting qubits[3], Boltzmann machines[4] or classical annealing[5] in memristor electronics, coherent Ising machines using a fiber-based optical parametric oscillator[6] or integrated optics[7,8] , as well as pure digital implementations using field programmable gate arrays (FPGA)[9] or graphics processing units (GPUs)[10]. These platforms emulate meta-heuristic algorithms that solve Quadratic Unconstrained Binary Optimization (QUBO) problems, with the ambition to more rapidly solve large problem sizes beyond those currently tackled by exact algorithms. Exact methods, such as branch-and-bound, are often limited to a few hundred variables per central processing unit (CPU) core for some typical QUBO instances[11,12] (see also Supplementary Material (SM) 1.1). At the algorithmic level, hybrid approaches have been proposed which allow subdivision of industrial-scale problems into QUBO-problems compatible with today's accelerator sizes[13,14] . Consequently, there is an incentive to find efficient accelerators compatible with this hybrid eco-system. In addition, despite contrary ambitions, quantum annealers have not been shown to provide speed-ups for industrial applications, while they bring high costs and complexity to deployment due to cryogenic cooling, adding a stronger motivation to leverage classical physics instead[6,9] .

Here we propose a new memristor-based annealing design that uses an energy efficient neuromorphic architecture instantiating a Hopfield Neural Network (HNN). An HNN is a fully-connected recurrent neural network without self-feedback, where a state of a neuron is dependent on the input received from all other neurons[15,16]. HNNs were initially proposed to implement associative memory for pattern recognition[15] and research on memristor HNNs have explored this type of application, along with other broad applications such as analog-digital converters[17], many of which only require a sparse connectivity[5,18–20] (connectivity being the degree of non-zero weights on the connection matrix). Here, we use the HNN model to solve computationally hard QUBO problems[21–23], due to its capability to dynamically find minima (global or local) of implicitly-defined cost-functions, referred to as the energy. Optimization problems are encoded in a matrix representing the set of objectives and/or constraints.

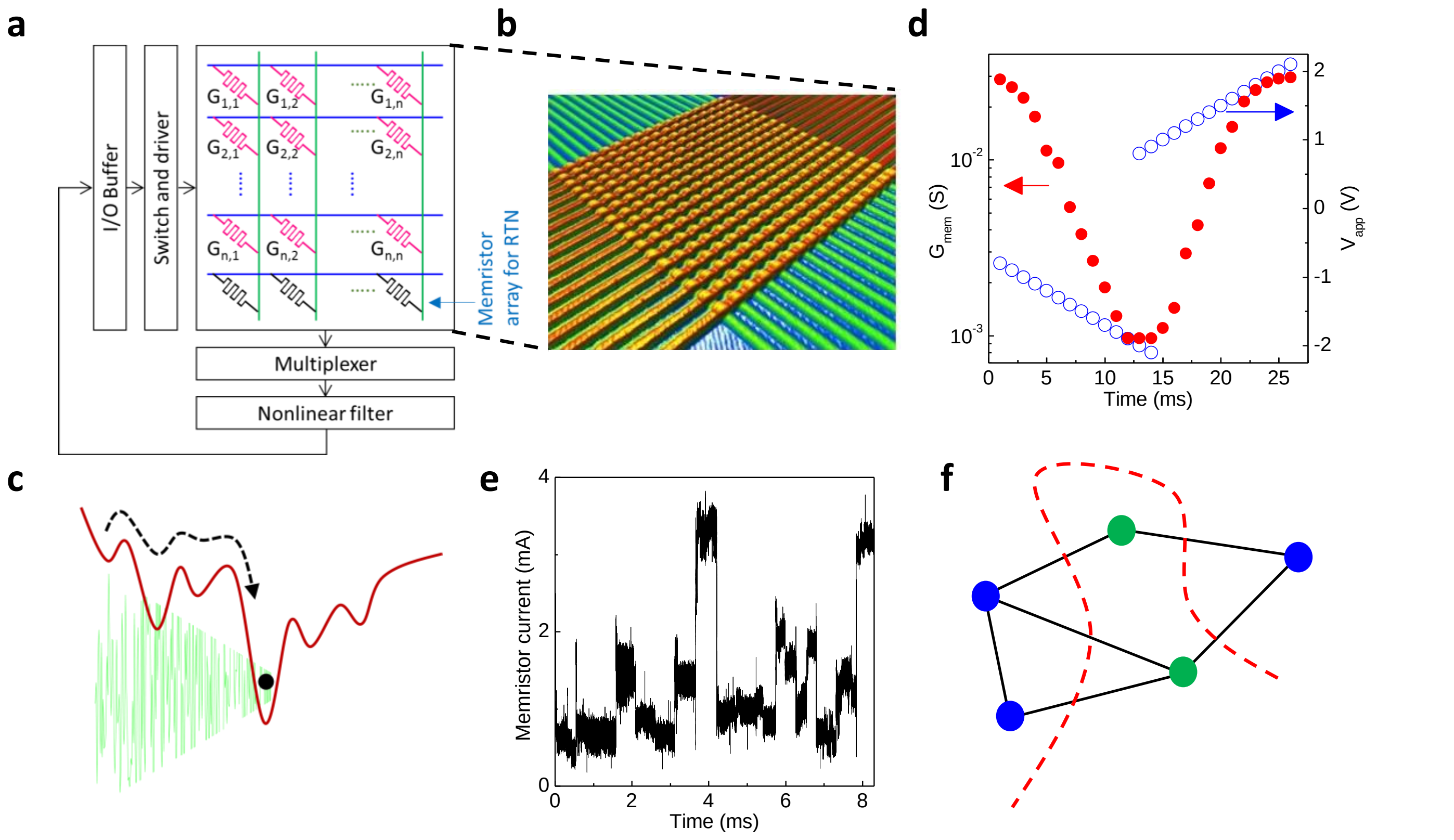

**Figure 1. Overview of the memristor Hopfield Neural Network (mem-HNN) and the Max-Cut problem.** (a) Schematic of the mem-HNN design including analog crossbar array and peripheral circuits (more details in SM 1.5) . (b) Atomic force micrograph of a typical memristor crossbar array. (c) Experimental data of electrical operation of a $TaO_x$ memristor exhibiting multi-stable conductance tuning via pulses. (d) Schematic energy landscape of a typical NP-hard problem, also illustrating simulated annealing. (e) Experimental data for random telegraph noise in a $TaO_x$ memristor. (f) A typical Max-Cut problem. Dashed red line cuts the graph into two sets of nodes, blue and green colored, such that the edges between the different colored nodes are maximized.

Our hybrid analogue/digital HNN architecture is mathematically equivalent to a conventional discrete-time binary HNN augmented with a noise contribution:

$$u_i = \sum_{j \neq i} W_{ij} v_j, \qquad v_i = \begin{cases} +1 & \text{if } (u_i + \eta_i) \geq \theta_i \\ -1 & \text{if } (u_i + \eta_i) < \theta_i \end{cases} \tag{1}$$

where $v$ is the state of the neuron, $u$ is the weighted feedback accumulated at the neurons at the next time step, $W$ is the zero-diagonal symmetric weights matrix, $\theta_i$ is a threshold and the vector $\eta$ represents the noise in the hardware nodes. The binary threshold function serves as an activation function, which essentially implements a nonlinear filter to process the recurrent weighted feedback.

The most power-intensive task here, namely vector-matrix multiplication, can be implemented using crossbars of memristors, particularly oxide-based two-terminal devices with tunable analog resistance levels[1], exploiting Ohm's and Kirchoff's law to obtain the vector-matrix product of input vectors and a stored weights matrix $W$ (Figures 1a-c). This enables highly efficient computations performed in the analog domain[24] and substantially reduces data movement as operations are performed directly in-memory[25,26] . The input vector is stored in the I/O buffer driving the rows of the array. To perform the vector-matrix multiplication, a switch matrix and driver circuitry, instead of a decoder, is required to enable all rows of the memristor crossbar in parallel. For a given cycle, a multiplexer is used to select one (or several) columns of the crossbar of memristors to calculate dot products for the neurons in a single operation. The output of the dot products directly feed into the nonlinear filters to perform the threshold function. The results are sent back to I/O buffers to update the binary status of the neurons, which are used as inputs for the next cycle (more details in SM 1.5). In this study, we will show that a memristor-based optimization network, a mem-HNN, is extremely fast, energy efficient, and can leverage intrinsic analog noise of the system to improve both solution quality and efficiency.

For NP-hard problems, even as problem sizes scale up only linearly, there are at least two main challenges with today's best known algorithms and processors: (a) resource consumption (time to completion, memory, etc.) grows exponentially and (b) an increasingly complex energy landscape evolves that can contain multiple local minima, a particular challenge when globally optimal solutions are desired (Figure 1d). When a noise-free ($\eta = 0$) discrete-time HNN is asynchronously updated (one node is chosen and updated at a time), the update rule of equation 1 assures that the HNN evolves in a way that only reduces the following energy function:

$$E = -\frac{1}{2}\sum_{i,j}^{N} W_{ij} v_i v_j + \sum_{i}^{N} \theta_i v_i \tag{2}$$

Through proper choice of the weight matrix $W$, an arbitrary optimization problem[21] can be encoded and solved with the HNN, which will eventually converge to a fixed point. However, noiseless discrete-time HNNs have not gained widespread adoption because, among other reasons, they could not achieve energy ascent to escape local minima, preventing them from solving large optimization problems with complex solution landscapes. Simulated annealing is a well-known technique to address this issue[27–29], wherein a stochastic or uncorrelated process is used to perturb the state of a system, thereby enabling it to achieve energy ascent sufficient to escape local minima. As the system approaches the global minimum, the magnitude of noise is tuned down to trap the system in the optimal solution (Figure 1d). Implementing controlled pseudo-random number generation to achieve simulated annealing typically requires elaborate digital circuits with unfavorable size, power-consumption, and latency properties[30]. In contrast, here we propose to utilize intrinsic analog noise from the crossbar circuit computations, supplemented by an extra row of memristors to inject a tunable level of random telegraph noise (RTN) to generate the needed independent noise (Figure 1a) in each column. This leads to a network model similar to a Boltzmann machine (see discussion in SM 1.4). Oxide memristors, such as those using $TaO_x$, have been shown capable of producing controlled random telegraph noise (RTN)[31,32] in certain regions of bias (Figure 1e). Another source for noise injection is the use of Mott memristors, which have been shown capable of tunable chaotic dynamics and have previously been proposed for implementing simulated annealing[33].

Given the diversity of existing and developing annealing approaches, there is a strong need for benchmarking [22]. We use the NP-hard graph problem, Max-Cut[21], as a benchmarking task, as it has already been extensively used in recent literature[6,7,9,10] . As illustrated in Figure 1f, in the Max-Cut problem the objective is to obtain a partitioning of the nodes of a connected graph, such that the number of connections between the nodes of the two partitions is maximized. This problem has relevant applications across industry, particularly in efficient circuit routing and minimizing the number of vias.

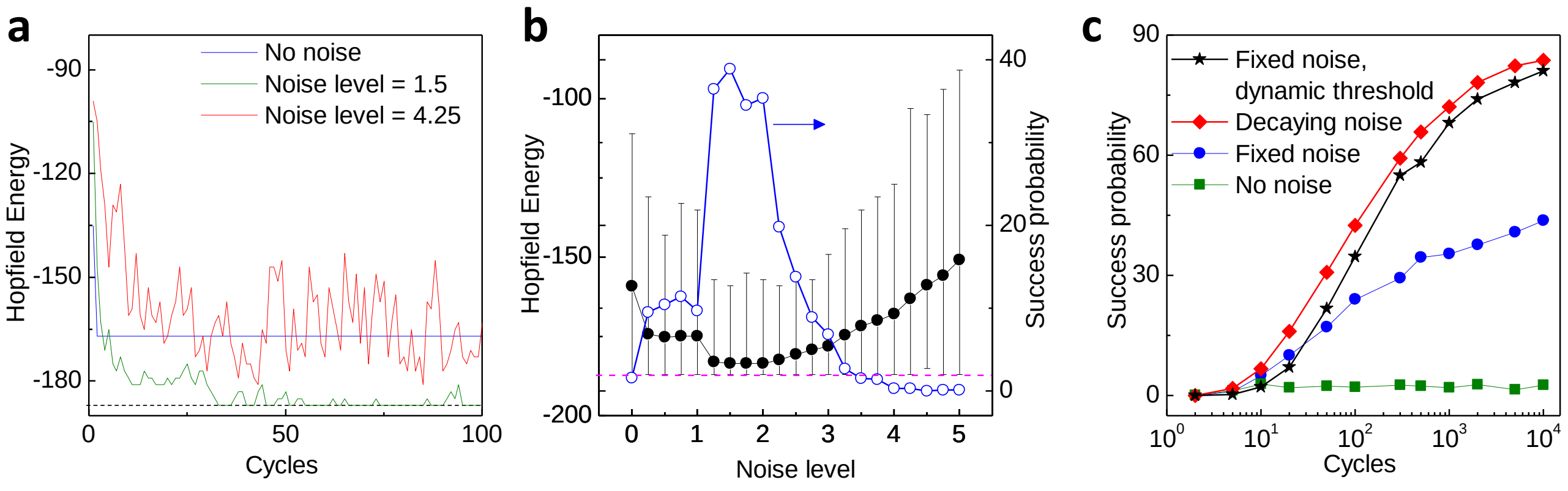

**Figure 2. Utility of noise in Hopfield Neural Networks to obtain better solutions, illustrated for 60-node instances of dense Max-Cut problems.** (a) Shows the influence of different noise levels on the outcome of the final stable state. Three different noise levels are shown: 0, 1.5, and 4.25. The dashed horizontal line is the energy corresponding to the optimal solution. (b) Mean and range of Max-Cut energy results with different fixed noise levels. At each noise level, 1000 instances with different random initial states are investigated. The error bars illustrate the maximum and minimum Max-Cut of all instances at the corresponding noise level. The success probability – the probability of achieving the known globally optimal solution – is also plotted versus noise level. (c) Comparison of the effect of using a quadratic superlinear decaying noise, a fixed noise, a fixed noise with a dynamic threshold, and no noise.

## Performance enhancement due to annealing schemes

The level of noise present in Equation 1 that governs our mem-HNN plays a vitally important role in the network's ability to find globally optimal solutions. As examples, Figure 2a investigates the effect of noise on the mem-HNN, starting with simulations at three different noise levels (0, 1.5 and 4.25) but otherwise using identical initial state conditions. We used unweighted Max-Cut instances provided in the Biq Mac Library for dense (50%) connectivity and node sizes of $N$ = 60 and larger [6,34] . As shown in Figure 2a, without any noise the Hopfield network converges to a sub-optimal solution within just a few cycles. This is because such dense graphs have a complex energy landscape and a conventional HNN can easily get stuck at local minima. Utilizing a larger noise level, say 4.25, it is very easy to get out of such local minima, overcoming barriers in the energy landscape, but the downside can be continued large fluctuations, even after discovering an optimal energy solution. Therefore, finding an appropriate noise level is a critical task.

To verify the existence of an optimal noise level, we ran a suite of simulations varying the noise levels (Figure 2b) and starting states, while tracking best, worst, and mean energy of solution. Shown in Figure 2b, by varying the noise level from 0 to 5, we can observe that the optimal level is around 1.5, which not only has the best mean value for Max-Cut energy, but also less variability and a peak in the probability of reaching the known globally optimal solution which we term here the ”success probability.” We also observe that as the noise level goes to zero (or well above 2), the success probability can tend to zero with the mean Max-Cut energy also dropping dramatically.

Based on our studies in Figures 2a-b, we know that noise is beneficial for escaping local minima, but converging to a final optimal solution is challenged if the network continues to fluctuate. To simultaneously harness the benefits of noise while improving network stability, a good approach[35] is to use a decaying noise profile which progressively reduces in magnitude in the final cycles so that the network may hold its state. This approach is leveraged in the simulated annealing algorithm, in which an initially high “temperature” for the system is slowly “cooled”[27]. Related approaches with modified annealing schedules have been used, particularly for increasing parallelism and reducing time spent in local minima[36].

To investigate the effects of the noise profile, we repeated the tests of Figure 2b with different cycle-dependent noise profiles (see also SM 1.3). In Figure 2c, we compare no noise, a fixed (constant) noise, a decaying noise, and a fixed noise with moving threshold versus increasing total cycles. In all cases, the amplitude of the noise has been separately optimized for best solution accuracy. The network showed the best performance with a simulated annealing approach where a quadratically decaying noise versus cycle number is applied (Figure S2). As noted earlier, the benefit here is that the system slowly becomes stabilized in an optimal solution. Our study shows that no noise performs the worst, while a fixed finite noise gives substantial improvements. Interestingly, comparable performance to the simulated annealing approach can be achieved even if it is not possible to tune down the noise profile. Instead, tuning the threshold function criteria ($\theta_i$ in Equation 1) to be increasingly strict for later cycles [9] performs nearly the same. This is an important result since implementation in analog crossbar arrays may not always allow noise to be tuned down. Instead, raising the threshold in the simple peripheral circuitry adds the desired stability to the Hopfield network in converging to fixed solutions, while harnessing the full noise in early cycles to climb out of local minima.

## Experimental implementation of in-memory analog computing in memristor crossbars

Figure 3a shows energy minimization results from an experimental realization of the mem-HNN system using a typical annealing schedule over 300 cycles. Figure 3b shows that it is experimentally possible to accelerate the annealing schedule over just 10 cycles and still achieve global minimization with the optimal amount of noise. A 60 node Max-Cut graph problem is programmed into a non-volatile memristor crossbar platform (chip image is shown in Fig. 3c) with the ability to activate and sense all array rows and columns simultaneously. The targeted and programmed weights matrices are shown in Figs. 3d and 3e, respectively. The core vector-matrix multiplication operations were performed within the memristor crossbar arrays, with peripheral sensing circuits implemented in printed circuit boards, and the nonlinear filtering and energy calculations performed within a control computer (detailed in Fig. S6) . Our mem-HNN performs the computations of Eq. 1 in the analog domain in a single clock cycle, offering high parallelism and power-efficient computations37. Additionally, we experimentally investigated the addition of a finite amount of noise, implemented here within the digital computer following the analog computation result, but before the nonlinear threshold function is applied. This allowed experimentation with no noise, a large fixed noise, and the optimal noise with a decaying profile determined in the previous section. In agreement with simulations (Figs. 2a and 4a), experiments show improved convergence to more optimal solutions with the injection of a finite amount of noise during the dynamical updates of the network. In addition to the utility of optimal noise, it was possible to achieve global minimization within just 10 cycles using an accelerated annealing scheme (Fig. 3b).

An important aspect manifested in the experimental platform is the potential to add a form of intrinsic analog noise in contrast to precise digital computing. This is due to non-idealities in any physical system, including distributions in the weight programming, parasitic resistances, capacitances, atomic fluctuations in nanoscale devices, and electronic trapping/de-trapping physics[31]. In previous work, we experimentally calibrated the sources of noise in the memristor analog computing platform[37] and determined it can be dominated by the interaction of finite wire resistances in the array with high conductance memristor cells as well as small errors in the programming of the memristors. We observe here (see Figure S6) and in previous work that these non-idealities lead to tight Gaussian-shaped error distributions around the ideal target result, and can be accurately captured in circuit-level simulations that include the parasitic resistances, finite ON/OFF ratio, and programming errors[37]. The variance of this distribution can grow with larger wire resistances within the rows and columns or simply growing the size of the array.

Figures 4a-b show circuit simulations of a 60×60 array using routinely manufacturable wire resistances and memristor parameters ("standard" values, corresponding to $R_{ON} = 10$ kΩ, $R_{OFF} = 1$ MΩ, $R_{wire} = 1$ Ω per block) as well as parameters substantially more non-ideal ("sub-standard", corresponding to $R_{ON} = 2$ kΩ, $R_{OFF} = 100$ kΩ, $R_{wire} = 1$Ω per block), both compared to fully ideal ("exact") results. A good match is seen between the ideal results and analog circuit results with standard parameters. It is even seen in Figure 4b that analog results can give slightly improved success probabilities over the exact results, and this is due to the addition of intrinsic analog noise which may have a more advantageous distribution than what is used in software. It is also evident that dramatically worse device and circuit parameters degrade the results, yet without a sharp drop. As highlighted in the above studies, intrinsic variability and non-idealities in analog arrays can be an advantageous resource, but too strong noise may compromise computational results.

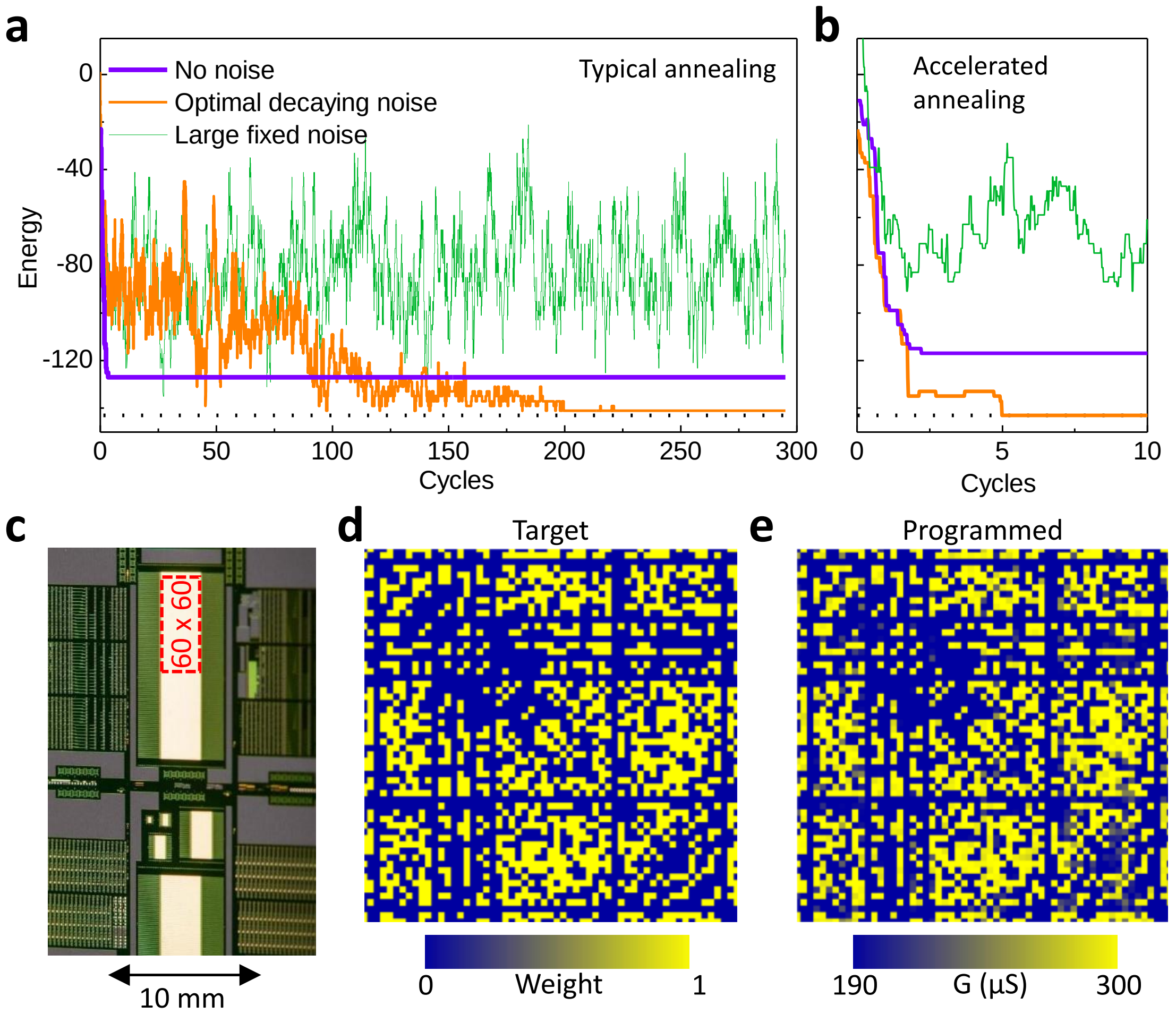

**Figure 3. Experimental implementation of the mem-HNN.** (a) shows experimental results of a 60 node Max-Cut problem solved in the analog domain in a memristor crossbar over 300 cycles. The energy of the solution found versus number of cycles is shown with no added noise, an optimal decaying noise, and a large fixed noise. The dashed horizontal line is the energy corresponding to the optimal solution. (b) is identical to (a), except in the total number of cycles used (10), demonstrating global optimization even with an annealing schedule accelerated relative to the typical case in (a). (c) Optical micrograph of the memristor crossbar chip used. A 60 × 60 sub-array within a 128 × 64 array was chosen for programming of the target weights matrix (red box). (d) Target weights matrix and (e) experimentally obtained conductance matrix after programming, showing close agreement.

These non-idealities can grow with array size as the additive wire resistances lead to large errors at the far edges of the array that also interact and are amplified by other non-linearities and non-idealities in the array[38]. To quantify the scalability limit of our mem-HNN system, we simulated a vast ensemble of analog computations performed in memristor crossbar arrays from a size of 8×8 up to 1024×1024 (Figure 4c). We tracked how the intrinsic noise in these analog arrays grows with size for both dense and sparse matrices. Higher density matrices (<50% non-zero values) lead to larger noise than sparse matrices (<10% non-zero values). In both cases, the analog computations in smaller arrays give results very close to ideal results in such a way that there is no overlap of neighboring integer distributions. In other words, the error distribution has a standard deviation that is sufficiently below 0.5 that any result can be rounded to the nearest integer and will match the exact integer (digital) result, see lower inset in Figure 4c for the noise distribution of a 32×32 array. Novel error-correction codes developed specifically for crossbar computations[39] can further assure perfect results with some redundancy overhead. On the other hand, larger arrays can develop errors of a magnitude that may no longer be tolerable or easily corrected, see upper inset to Figure 4c for a 128×128 array. It is seen in Figure 4c that digitally-equivalent precise computations may be computed in analog arrays of sizes up to 100×100 for dense graphs, and up to 256×256 for sparse graphs. Arrays larger than this will add a finite amount of noise to the analog computations. We emphasize, however, that this does not preclude the use of the mem-HNN system for solving larger graph problems, but does require the sub-tiling of larger graph problems across multiple smaller arrays, very similar to what has been architected in machine learning accelerators with memristor analog systems[25,40]. However, an important observation here is that in solving optimization graph problems that grow in size, these require increasing amounts of injected noise during computations, as the size of the energy barriers also typically scales. This is captured in the Figure 4c plot of the "optimal noise" needed, where simulations across an ensemble of different dense graph problems of varying sizes were solved and the optimal amplitude of noise to achieve the best success probabilities was computed. This shows a favorable match between the need for intrinsic noise to accelerate and improve the solution of optimization problems and the capability of analog arrays to supply this amount of noise up to fairly large arrays. As shown, even arrays up to a size of 1024×1024 generate less analog noise than that which is optimally required. This enables extremely large arrays to be utilized with supplemental noise injected using our mem-HNN scheme of Figure 1a. Large arrays are highly desirable from an area and power perspective as it allows the expensive peripheral circuitry to be amortized across many more parallel computations[38] performed. Thus, the performance gains for analog-based optimization problem solvers may be expected to be even greater than that of analog-based machine learning accelerators[25,41].

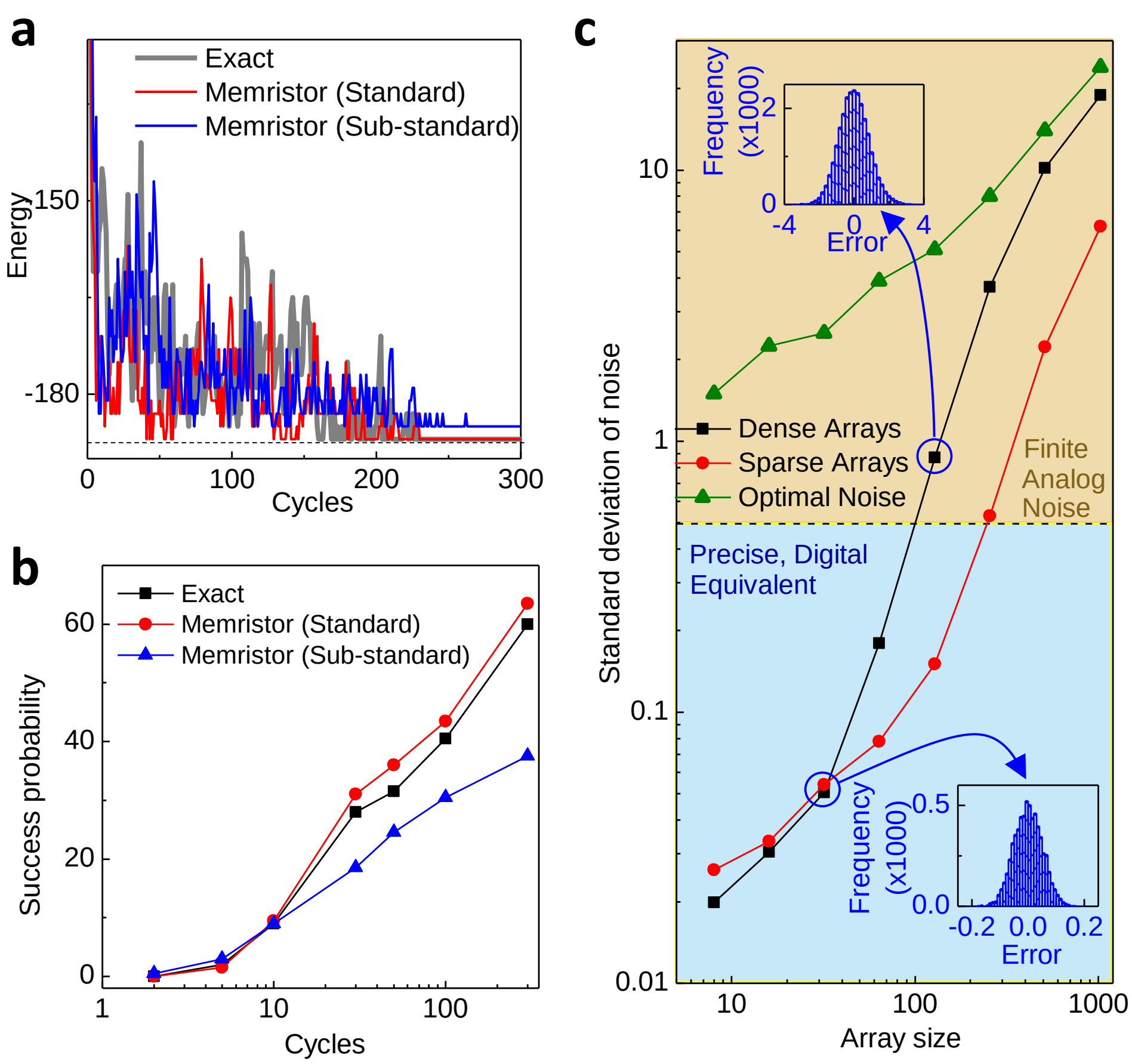

**Figure 4. Circuit and array simulations of the mem-HNN.** (a) Circuit simulations of the mem-HNN, which includes both the exact (digital software) results along with circuit simulations using routine ("Standard") memristor parameters and poor ones ("Sub-standard"). The dashed horizontal line is the energy corresponding to the optimal solution. It is seen in (b) that the overall success probability with analog memristors closely matches the exact results, although poor memristor parameters can degrade the computations. As array size grows, increasing non-idealities can lead to noisy computations that no longer match the exact ideal results. (c) The scaling of errors (standard deviation of noise) with array size (the number of components in the array being the square of the array size) for dense and sparse graphs, along with the amount of optimal noise needed in solving graph problems. It is seen that the intrinsic noise in analog arrays is still below that needed up to arrays of 1024 × 1024, giving a favorable scalability of the mem-HNN. Insets to (c) show the error distributions corresponding to data points indicated by blue arrows for 32 × 32 and 128 × 128.

## Mem-HNN Performance and Scaling with Problem Size

Having validated our mem-HNN system with experimental measurements, we were able to explore and benchmark our system in simulation for problem sizes and time-scales beyond what has been experimentally feasible to date. Figure 5 explores the problem size scaling, speed, and energy consumption in solving dense Max-Cut problems (connectivity of 50%) of increasing size. All studies leverage noise to enhance solution quality and number of cycles (as shown in Figure 2). Figure 5a shows that for increasing problem sizes (graphs) it becomes more difficult to find the globally optimal solution, reducing the success probability, but increasing the number of cycles can mitigate this effect.

Figure 5b shows the computation time of the mem-HNN to reach a 99% probability of optimal solution [9] for increasing sizes of dense graphs. As illustrated in Figure1, larger problem sizes require larger memristor crossbars and peripheral circuitry. Due to the serial update of each node, the latency scales linearly with the problem size. However, batches of nodes can be updated in parallel with a small impact on the solution probability but a net reduction in time-to-solution.

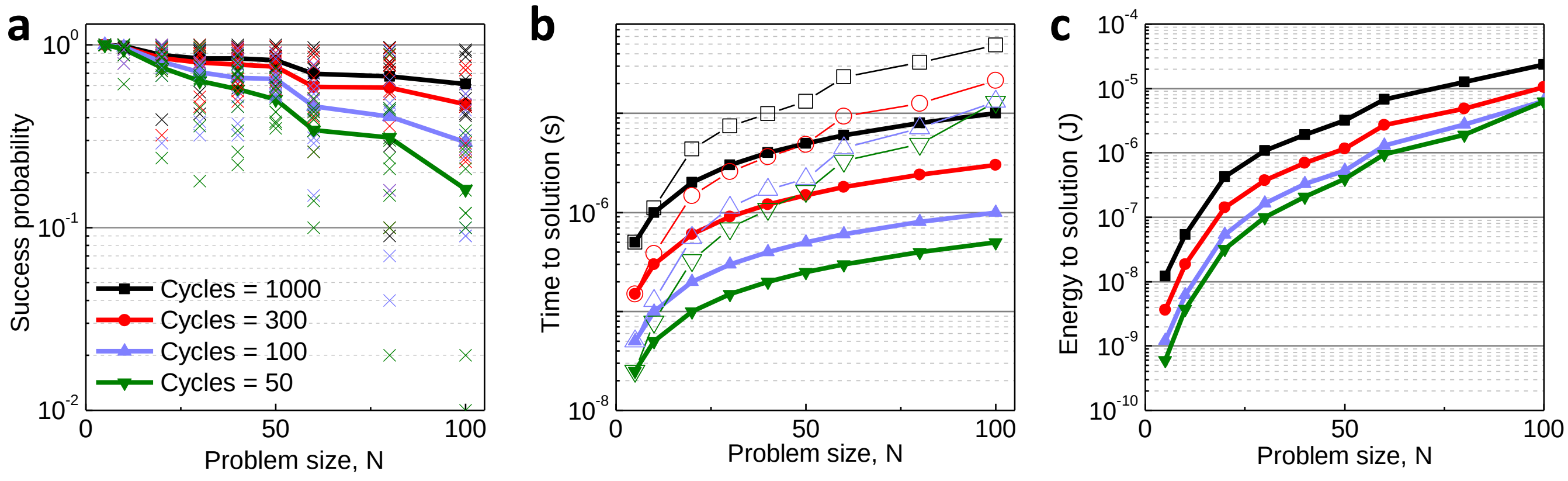

**Figure 5. Simulations of the memristor-Hopfield Neural Network solving dense instances of the Max-Cut problem with varying graph sizes. (**a) Probability to reach the globally optimal solution versus problem size. For each problem size, ten separate graph instances are computed and the total number of cycles used is varied from 50 to 1000. (b) Shows the time-to-solution with 99% probability of success. Memristor crossbars can either be utilized in sequential runs, operating multiple times with differing starting states to ensure 99% success (open symbols). Or, multiple crossbars can run in parallel (filled symbols) at the cost of larger area. (c) Energy to reach 99% optimal solutions is computed as a function of problem size. Larger problems require larger memristor crossbars, more peripheral circuitry, and also more parallel crossbars to ensure a 99% success probability is attained. The total energy consumed is not affected by the choice of parallel or sequential runs.

Figure 5 uses batches of ten nodes at once to speed up computation parallelism, with future improvements possible with more optimal schemes[36]. Additionally, instead of running the same problem instance sequentially on the same crossbar, problems can be solved on multiple crossbars in parallel, with variable initial states, in order to reach the target 99% probability. In the latter case, the time-to-solution scales only with problem size as shown in Figure 5b (filled symbols).

Energy consumption for the mem-HNN is shown in Figure 5c. This plot captures energy consumed in the peripheral Mux, decoders, comparators, and the analog crossbars themselves. Additionally, larger graph problems have a reduced probability of solution (Figure 5a), thus requiring increasing number of parallel mem-HNN units. Nonetheless, the efficiency of performing the needed multiply and add operations (Equation 1) of a Hopfield network in the analog domain using memristor crossbars is shown here. Our system (see next section) outperforms current digital systems (CPUs and GPUs) by approximately 10,000, with even larger gains compared to current quantum annealers.

## Comparison between mem-HNN and other annealing accelerators

In Figure 5, we demonstrated how the mem-HNN, due to its ability to have all-to-all connectivity, seems to have a state-of-the-art scaling of success probability, time to solution and energy to solution for the problem sizes considered in this paper. In Table 1, we compare the performance of our proposed mem-HNN with other technologies in solving dense Max-Cut problems with 50% connectivity and graph size $N = 60$. These values are based on experimentally-grounded simulations for the mem-HNN, whereas fully experimental results for the other technologies are taken from literature. We list some relevant metrics which affect the throughput of the system (connectivity, annealing time, time-to-solution), the energy efficiency (power, energy to solution), and potential for scalability (connectivity, scalability of the success probability $p_{rs}$).

In this table, we track four different types of technologies: a mem-HNN, a fiber-based coherent Ising machine (CIM) with an FPGA in the feedback loop[6], a state-of-the-art implementation of parallel tempering (PT) on a single Intel (R) Core (TM) i7-3720QM CPU @ 2.60GHz (details in SM 1.8.3), D-wave's 2000Q quantum annealer (containing 2048 qubits)[6] and an NVIDIA GeForce GTX 1080 Ti GPU running a noisy mean field annealing (NMFA) algorithm[10]. The mem-HNN system utilizes analogue computing with nano-electronics, the CIM system is a hybrid electronic and optical accelerator, while the GPU and CPU platforms represent digital hardware running physics-inspired heuristics.

Previous comparisons between emerging annealing technologies have primarily focused on the run-time required to reach a solution with high likelihood (time-to-solution). In these studies, the energy efficiency typically received less attention (as noted before[42]). As energy-efficiency is a critical metric in a post-Dennard computing world, we compared this for the listed technologies based on estimated power consumption. Note that both the D-wave and CIM system are proof-of-concept systems which are not currently optimized for energy-efficiency (SM 1.8.2). In contrast, energy efficiencies for our mem-HNN are readily estimated based on extensive studies of memristor crossbar arrays for matrix-vector multiplication applications[37,41,43]. While we do not have access to the power consumption of the GPU during implementation of the noisy mean field algorithm, we use the available power specs for this model as a realistic upper estimate. For the CPU, an effective power consumption of 20 W has been estimated[42,44].

**Table 1.** Comparison of the mem-HNN and current state-of-the-art annealing accelerators, such as a GPU implementation of the noisy mean-field algorithm[10], our own simulations using the previously suggested[6] parallel tempering implementation on a CPU (cfr. SM 1.8.3) and experimental results for the D-wave annealer and the measurement-feedback CIM discussed in Ref[6]. A hybrid update mechanism updates some, but not all the nodes at a given iteration.

| | **mem-HNN (seq.) memristor** | **mem-HNN (par.) memristor** | **NMFA GPU** | **PT@UFO single-core CPU** | **D-wave 2000Q supercond. qubits** | **CIM fiber-optics** |
|---|---|---|---|---|---|---|
| **Clock frequency** | 1 GHz | 1 GHz | 1.582 GHz | 2.6 GHz | | 1 GHz |
| **Annealing time $T_{ann}$** | 300 ns | 300 ns | 12.3 $\mu$s ($N = 100$) | 223.6 $\mu$s | 1 ms (N=55) | 150 $\mu$s |
| **Time-to-solution** | 3.3 $\mu$s | 0.3 $\mu$s | 10 $\mu$s | 223.6 $\mu$s | $10^4$ s (N=55) | 600 $\mu$s |
| **Power** | 66 mW | 792 mW | < 250 W | 20 W | 25 kW | |
| **Energy-to-solution** | 0.22 $\mu$J | 0.22 $\mu$J | <2.5 mJ | 4 mJ | 250 MJ | |
| **Solutions/s/Watts** | $4.6 \times 10^6$ | $4.6 \times 10^6$ | > 400 | 250 | $4 \times 10^{-9}$ | |
| **Update mechanism** | hybrid | hybrid | asynchronous | asynchronous | synchronous | asynchronous |
| **Connectivity** | all-to-all | all-to-all | all-to-all | all-to-all | Chimera | all-to-all |
| **Scaling $p_{rs}$** | $ae^{-bN}$ | $ae^{-bN}$ | $ae^{-bN}$ | $ae^{-bN}$ | $ae^{-bN^2}$ | $ae^{-bN}$ |
| **Cryogenic cooling** | no | no | no | no | yes | no |

Overall, it is evident that electronics-based approaches currently outperform both the quantum annealing and the optical systems (see SM 1.8 for a more detailed discussion). Yet even among the electronic solutions, the mem-HNN substantially outperforms both the GPU and CPU in terms of time-to-solution (by at least 10-fold), but even more so in terms of energy to solution (by at least 10,000-fold). These performance gains can be attributed to (1) the highly efficient matrix vector multiplication performed in the analog domain, (2) intrinsic access to a tunable noise source within the memristor crossbar circuit, (3) high parallelism in the dot-product operations as well as multiple crossbar circuits running separate instances of the graph problem with differing starting states to speed-up solution convergence (an approach also harnessed across GPU cores), and (4) a substantial reduction of data fetching and communication through the use of in-memory computing[45,46]. The latter feature refers to the elimination of the well-known von Neumann bottleneck where computation speed and energy consumption is dominated by bringing data from local memory or storage.

Fundamentally, the D-wave system is the only system without all-to-all connectivity between its nodes, which leads to costly overhead in the required number of Ising spins to implement dense connectivity matrices. Consequently, a poor scaling as a function of problem size has been observed for dense problems[6]. Note that even for a relatively small $N = 60$ problem, D-wave's metrics are unfavorable compared to the other technologies. For these tasks, therefore, the benefits due to D-wave's potential use of quantum effects are negligible.

The mem-HNN has similarities in approach to the acceleration of machine learning operations in more modern neural networks using memristor crossbars[37,43,47,48] . An important difference is that the present system need not act as a fully-precise digital replacement[25], which would come at a high performance cost. Instead, the mem-HNN shows even higher performance potential over fully digital systems (by 10,000-fold) by operating deeper within the analog domain with desirable intrinsic noise. We further suggest that there is ample room for future improvements. The use of continuously tunable resistive cells allows our mem-HNN to solve graph problems requiring multiple bit precision in the connection matrices, without raising the power consumption. Notably, over 6-bit resolution has been experimentally shown[37], while architectures using tiling for arbitrarily high bit precision have been previously described[25]. Future circuits and architectures that leverage other low-power analog elements such as negative-differential resistance (NDR)[49] could eliminate nearly all digital functions (e.g., the nonlinear filter of Figure 1a) present in the current design. Un-clocked asynchronous designs have further potential for performance gains, but will require deeper co-design of the algorithm with the physical hardware implementation. Finally, we note that in order for the present design to support arbitrarily large problem sizes that are well-beyond single crossbars, a careful hierarchical architecture analogous to that in memristor-based machine learning systems[4,25] is required. A future research direction is to assess how the introduction of hierarchy and tiling in such a combinatorial accelerator will affect the time and energy scaling. The present study points to a very competitive approach to solving computationally intractable problems by utilizing analogue-domain memristor crossbar computations. The mem-HNN system is a hybrid analogue/digital computing platform that will benefit from future CMOS technology nodes due to compatible semiconductor processing. Memristors have shown size scalability[50] to 2 nm and fast operations[51] well below 1 ns. Augmenting digital technology with increasingly energy-efficient analogue components further opens up analogue-scaling that have the potential to continue even after the end of Moore's law, providing future technological benefits not found elsewhere.

## Acknowledgements

We are grateful to Salvatore Mandra for performing the CPU simulations used in Table 1 and early review of the manuscript. We acknowledge helpful discussions with Helmut Katzgraber, Peter L. McMahon, Edward Rothberg, Karl Roenigk, Cipriano Santos, Richart Slusher, and Jeffrey Weinschenk. This research was based upon work supported by the Office of the Director of National Intelligence (ODNI), Intelligence Advanced Research Projects Activity (IARPA), via contract number 2017-17013000002.

## Supplementary Material

### 1.1 Problem landscape

Good examples of the enterprise relevant NP-hard problems that future combinatorial optimization accelerators should target to solve can be found in the benchmark sets provided by the Mixed Integer Programming (MIP) community. Whereas we mentioned in our introduction that Ref.[12] demonstrated that some typical papers discussing the solution of QUBO's using exact algorithms were focusing typically on problem sizes of a few 100 variables on a single CPU, it should be noted that plenty of larger-scale MIP instances are known to be solvable using an out-of-the-box solver on a standard desktop in even less than one hour (cfr. the 'easy' category in MIPLIB2017[52]). However, the latter, 'easier' problems do not necessarily have a quadratic term (and are hence not a QUBO), are relatively sparse, or have another problem-specific feature in which the usage of the extensive range of exact methods incorporated in software frameworks such as Gurobi or CPLEX can be beneficial. The Max-Cut problem instances discussed in this paper, tend to be not easily solvable (using the MIPLIB-definition of easy) on a single CPU for problems of system size $N$ = 60,100. Nevertheless, it is important to acknowledge that when mapping problem instances in the MIP-format to the QUBO-formalism that corresponds with our annealing accelerator, we might, for some problem instances, miss opportunities to use problem-specific tricks. This is one of the reasons why we believe that future architectures relying on annealing accelerators, should use hybrid algorithms that contain some form of pre-processing and can identify whether running either the entire problem or some of its sub-problems on accelerator hardware is justified or not, by potentially even allowing to run analogue accelerators, digital meta-heuristics and/or exact techniques in parallel where appropriate.

### 1.2 Mapping the Max-Cut Problem to Hopfield Network

Maximum cut or the Max-Cut problem is a classic NP-hard problem which finds a maximum cut in a graph[21]. In the Max-Cut problem, a partition of $S$ of all the vertices $V$ needs to be determined such that the number of edges (which are referred to as the "Cut") between $S$ and its complementary subset is maximized. In a graph $G(V,E)$, a cut vector can be defined as:

$$x_i = \begin{cases} +1 & i \in S \\ -1 & i \in V \backslash S \end{cases} \tag{3}$$

where all the vertices in S are represented by +1, and all the vertices in its complementary set are noted as -1. Based on references[21,53], we derive the Hamiltonian for the Max-Cut problem starting from the following figure of merit:

$$F.O.M. = \max \sum_{ij \in \sigma(S)} a_{ij} = \max \sum_{i<j} a_{ij} \frac{1 - x_i x_j}{2}, \quad \sigma(S) = ij \in E : i \in S, j \in V \backslash S \tag{2}$$

where $a_{ij}$ is the weight of the so-called Adjacency Matrix $A$.

This optimization problem can be transformed to a Hopfield energy function as follows:

$$\begin{aligned} E &= \min \sum_{i<j} a_{ij} \frac{1 - x_i x_j}{2} \\ &= \min \frac{1}{2} \left( \sum_{i<j} a_{ij} x_i x_j - \sum_{i<j} a_{ij} \right) \\ &= \min \frac{1}{2} \sum_{i<j} a_{ij} x_i x_j - C \\ &= \min \frac{1}{2} \sum_{i<j} a_{ij} x_i x_j \end{aligned}$$

$$= \min -\frac{1}{2}\sum_{i<j}(-a_{ij})x_i x_j \tag{5}$$

Consequently, the Max-Cut problem can be solved by minimizing the following Hopfield network energy function:

$$E = -\frac{1}{2}\sum_{i<j} w_{ij}x_i x_j, \quad w_{ij} = -a_{ij} \tag{6}$$

and the weight matrix of the Hopfield network is the negative Adjacency matrix

$$w_{ij} = -a_{ij} \tag{7}$$

where the number of Max-Cut can be derived by:

$$Max\ Cut = -\frac{1}{2}\sum_{i<j} w_{ij} - E \tag{8}$$

### 1.3 Investigation of the effect of different noise profiles

In this section, we compare the effect of using different noise functions in HNNs and explain why we choose a quadratic superlinear decaying profile as our default noise decaying function in the main manuscript.

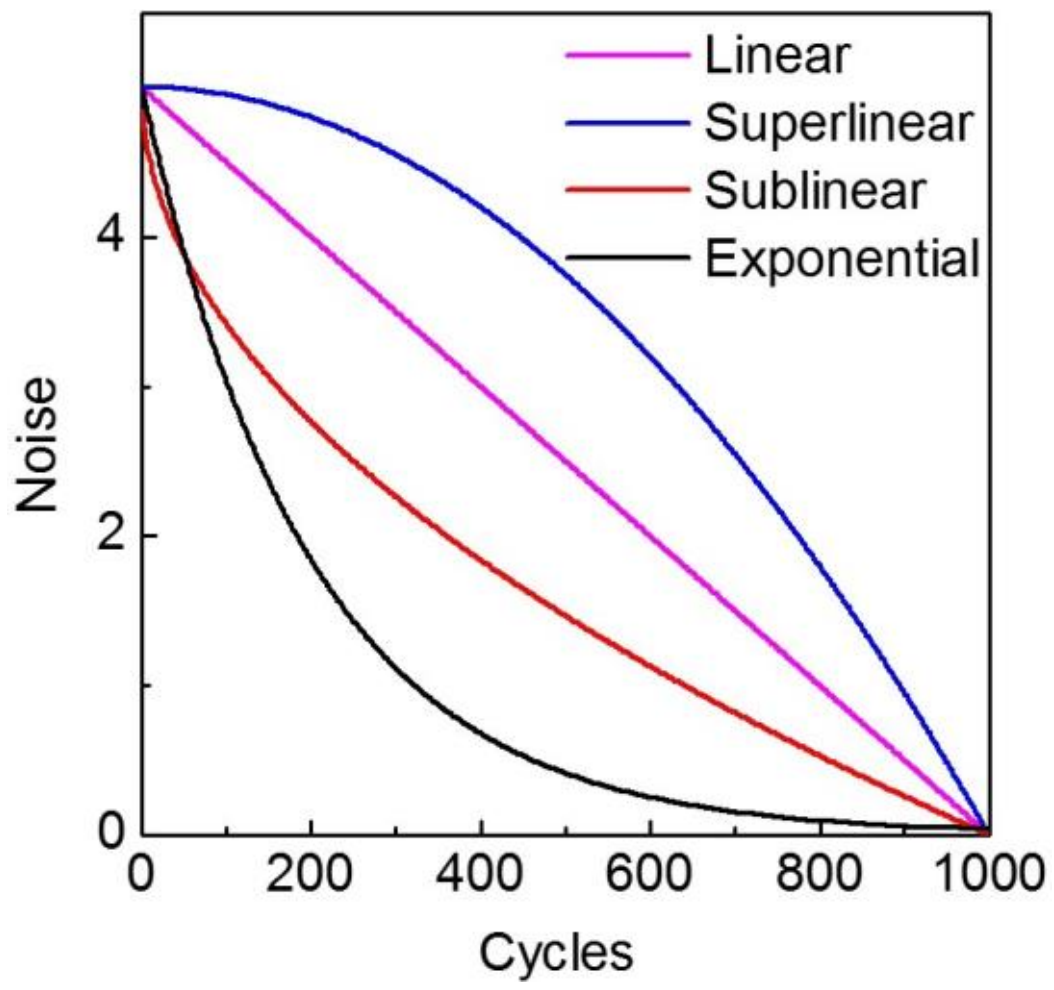

**Figure S1.** Four different decaying profile illustrated for a given annealing time $T = 1000$, and initial noise level 5.

For this purpose, we choose four qualitatively different decaying profiles for our random noise function: linear decay, quadratic superlinear decay, quadratic sublinear decay and exponential decay, to optimize the effect of injecting noise.

The mathematical expressions of the four different decay profiles are shown in Figure S1. These four decaying noise functions are also compared with the fixed noise function used in Figure 2b. In this comparison, all the initial noise levels of the four decaying profiles are set to 5, which we identified to be the optimal initial noise level for decaying noise of the 60 node Max-Cut problem. The comparison result is shown in Figure S2

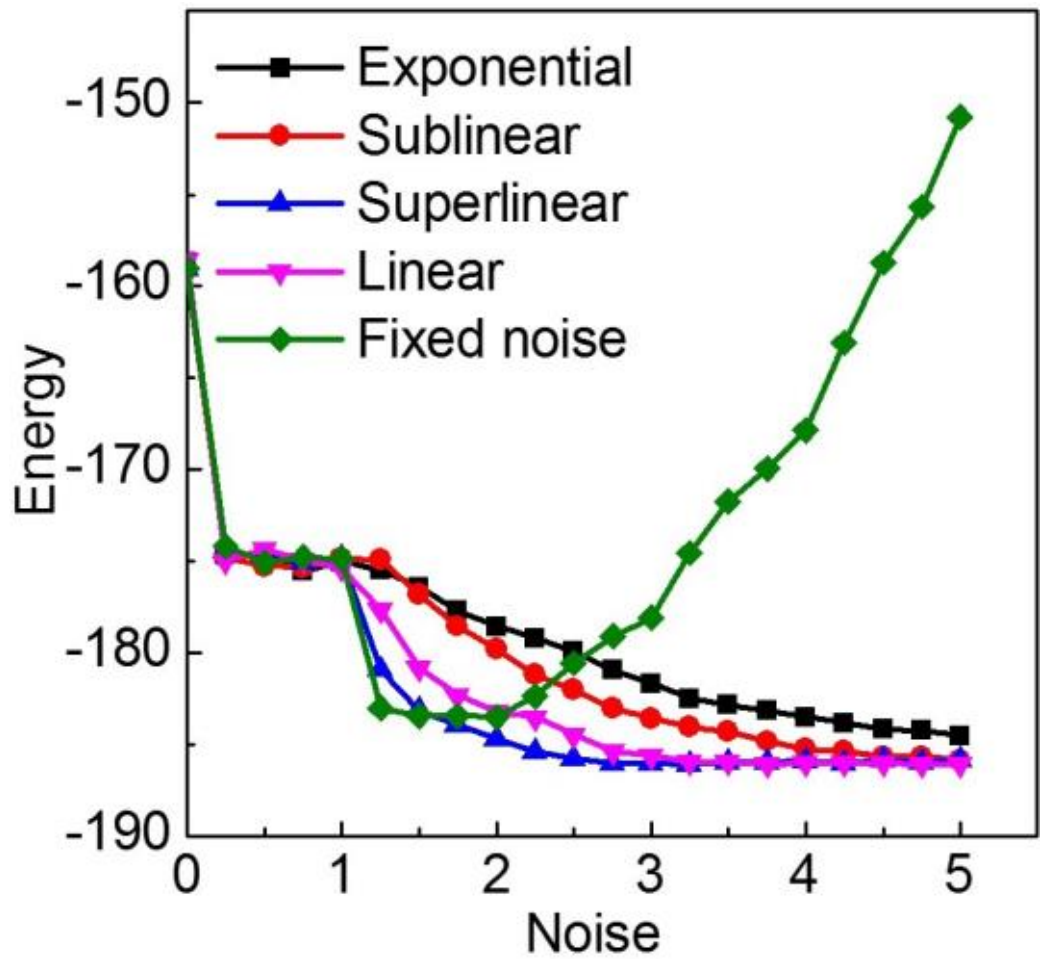

**Figure S2.** We compare five different noise profiles: four different decaying noise profiles and one fixed noise profile. The initial noise level of all four decaying profiles is set to be 5. For this benchmark task, the quadratic superlinear decaying profile leads to the best energy result.

Therefore, unless mentioned otherwise, to obtain the best energy performance when injecting noise, we select quadratic superlinear decaying noise in the HNN implementation in the main manuscript.

### 1.4 Function of the additive noise term in the stochastic mem-HNN

In this section, we first give an overview of different ways noise has been deployed in HNNs, followed by some context related to the specific design choice made in this paper.

#### *1.4.1 Historic overview*

The HNN has originally been proposed as an asynchronously updated discrete-time system with binary nodes without explicit noise terms[15]. Whereas it can be proven that this system will eventually converge to a stable fixed point[54], which fixpoint it relaxes into is both update scheme and initial condition dependent, and due to the limited set of neighboring states for a given state, it is easy for the network to get stuck in a local minima. Consequently, different versions of continuous-time HNN have been proposed, which should theoretically reach the global minima in a more efficient way than discrete-time HNNs, but suffer, amongst other things, from errors induced by the need to discretize the equations to solve them in software[55,56]. Using noise as a resource to avoid local minima was proposed using schemes that provide the theoretically optimal noise distribution[57], which is either computationally intensive or challenging to obtain in hardware. In other work, bounds on the noise have been derived that determine under which noise conditions the HNN will still converge to the desired limit set[58]. Boltzmann machines do not have explicit noise terms, but they are, similarly to Hopfield networks, discrete time systems and its updates are determined by asynchronously sampling the nodes of the network one-by-one following a sigmoid activation function[59]. Whereas Boltzmann machines have multiple applications, combining this system with simulated annealing allows to find ground states of the QUBO-system which is encoded in its Hamiltonian. As an alternative to this stochastic behavior, the mean-field annealing algorithm has been proposed, in which the output of the nodes is again determinstic, but instead of binary it is now real-valued, and follows the previously mentioned sigmoid activation function, of which the strength of the slope at the origin is adiabatically increased until it approximates the original binary heaviside (or, as in this work, a sign function when using a *tanh* as activation function with bipolar node states). A recent addition to that algorithm was to also explicitly include noise terms in the update equations, making the model again stochastic. This is called the noisy mean-field algorithm[10]. The latter approach is physics-inspired, based on the CIM results by Ref.[6], but so far, to the best of our knowledge, no published study has explicitly quantified how much this additional noise improves the solution quality compared to the original mean-field annealing algorithm. The noisy mean-field model was mainly proposed in a phenomenological way to explain the success probability scaling as a function of problem size for the CIM results.

In the world of physics-based hardware accelerators, the importance of noise has been discussed multiple times: recent proposals also show the benefit of having either classical or quantum fluctuations to enhance the chance to find the optimal solutions of NP-hard problems in Hopfield or Ising-based systems[6,7,60–65] . This is in line with other assessments that noise in electronics can be useful[66], as it plays an important role in some neural information processing schemes[67,68] . Based on the success of recent progress in physics-based accelerators, new software-algorithms for combinatorial optimization have been developed, sometimes explicitly based on the phenomenological behavior of the aforementioned physics-based accelerators, and these have been deployed in 'traditional' digital accelerators like GPUs[10] or FPGAs[9] - these algorithms all have an important stochastic or noise-induced component.

#### *1.4.2 Link between the stochastic mem-HNN and a Boltzmann machine*

In the theoretic simulations in this paper in, e.g., Figure 2, we are using decaying additive uniform noise in a discrete-time Hopfield Neural Network, an idea which has been proposed in the context of discrete-time HNNs for solving the shortest path problem[35], and some design criteria have theoretically been proposed in Ref.[69]. Neither of these two papers discloses that having a discrete-time HNN with binary states and additive uniform white noise can be seen as an approximated version of a Boltzmann machine, where the sigmoid function, i.e., the function typically used as the stochastic activation function, is now replaced by a (three-piece) piecewise linear approximation. In this paper, as intuitively expected, when performing simulated annealing by having the amplitude of the noise term gradually decay as a function of time, this can be interpreted as an approximation of a corresponding SA process in a traditional Boltzmann machine. Using such coarse piecewise linear approximations of the sigmoid function has been previously proposed to simplify the hardware implementation of a Boltzmann machine in an FPGA[70], but in that case the piecewise linear function was explicitly constructed in hardware, resulting in avoidable circuit overhead. In contrast, in our system we avoid the need to explicitly implement the piecewise linear system and restrict the hardware requirements to a simple comparator. More elaborate piecewise linear approximations have been successfully used in the context of FPGA-implementations of restricted Boltzmann machines for deep learning applications where the convergence degradation is claimed to be negligible[71,72] . One way to let the acceptance or rejection of a stochastic bit follow an activation function which follows the sigmoid more closely is by having a noise distribution which is Gaussian or a *sech* (the latter option being exact).

One recent implementation of simulated annealing in a mem-HNN proposed the use of hardware-generated noise in Cu-based CBRAM devices, as these devices have a tunable stochasticity[5]. In that scenario, a scheme similar to the one implemented in Fujutsi's digital annealer (DA)[9] was used in the sense that one spin (or a small set of spins) was selected at a time, to calculate the energy-difference $\Delta H$ which would be the result of a spin-flip, and this energy difference was then used to set the probability of the Cu-based CBRAM device by having an input pulse with length proportional to $\Delta H$. Whereas it is clearly beneficial to use hardware as an intrinsic noise source over the usage of RNG implemented in digital electronics, the specific implementation is rather slow (*ms* timescale), and contains additional circuit complexity compared to the one presented in this paper.

### 1.5 Design, area, and power for crossbar mem-HNN

When we employ a memristor crossbar to implement the vector-matrix multiplication, there can be different polarity options for both the input/output vectors and the matrix. They can have all positive or negative values (namely unipolar), or have partly positive and negative values (namely bipolar). In order to implement a bipolar vector and/or matrix, we can map the positive values and negative values to different memristor crossbars. Then, we can combine the calculation results from different crossbars with an extra circuit. For example, if we only have a bipolar vector or only a bipolar matrix, we need at least two memristor crossbars to complete the vector-matrix multiplication; however, if we have both a bipolar vector and a bipolar matrix, we need at least four memristor crossbars. In order to combine the intermediate results from each crossbar, we have to design additional circuits, thus resulting in complexity in circuit design. In this paper, we propose a general memristor crossbar circuit design, which is able to implement both unipolar or bipolar vectors and matrices in only one memristor crossbar. In our design, we use 4 memristor cells from two adjacent rows and columns, as shown in Figure S3a, as a unit to implement a bitwise multiplication. The two rows are always complementary and the diagonal memristor cells always see the same conductance value. Figure S3b shows the mapping scheme of one bipolar operand from the vector as the voltages applied to wordlines (WLs) and another operand from the matrix as the conductance of 4 memristor cells. Figure S3c gives two examples showing how the bitwise multiplication is performed with the proposed 4 1T1M unit design via Kirchoff's law. The output is essentially the difference between I and Ib, implemented by a comparator.

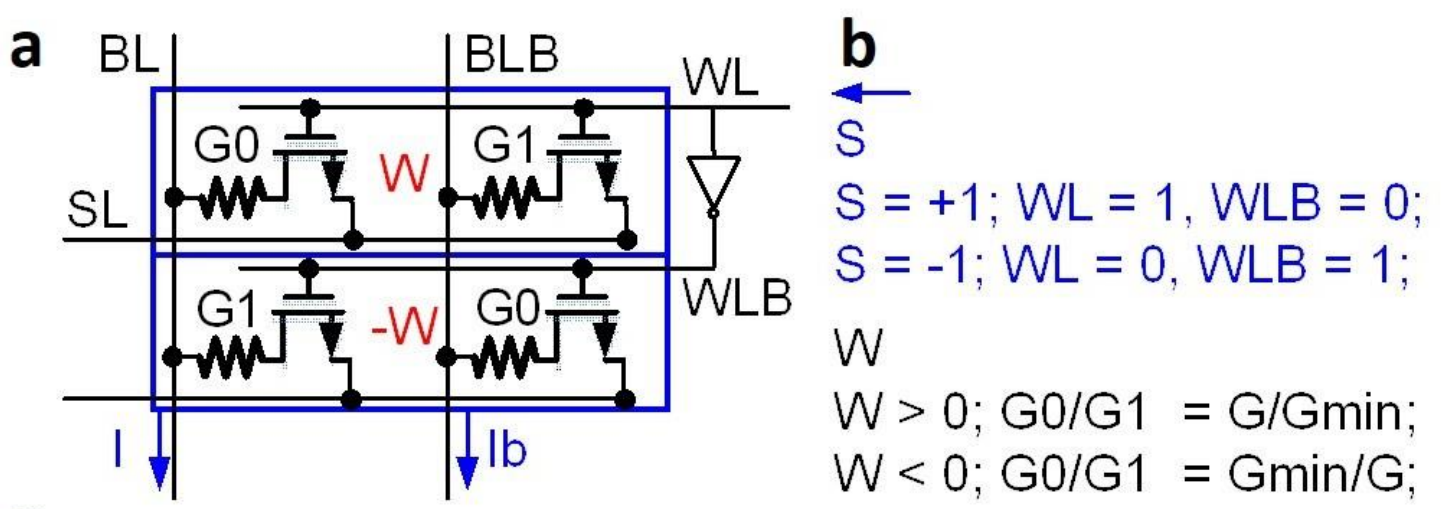

c

| S | WL | WLB | W | G0 | G1 | Value | I vs Ib |
|---|---|---|---|---|---|---|---|
| +1 | 1 | 0 | W > 0 | G | Gmin | > 0 | I > Ib |
| -1 | 0 | 1 | | | | < 0 | I < Ib |

**Figure S3.** Implementation of the bitwise multiplication of a bipolar vector and matrix using 4 memristor cells from two rows and two columns. (a) Circuit schematic of 4 1-transistor-1-memristor (1T1M) design. (b) The illustration of implementation of a bipolar vector value and a bipolar matrix value with 4 1T1M unit. (c) An example of bitwise multiplication as the output column current via Kirchoff's law.

With the proposed circuit unit, Figure S4 shows the circuit diagram of a 128×128 memristor crossbar for a Hopfield Neural Network:

1. An I/O buffer is designed to store the input vector, which is 64-bit and each bit is converted to two complementary voltages to drive two rows or WLs of the 1T1M crossbar;
2. A switch matrix and driver is able to enable multiple rows of the 1T1M crossbar;
3. At the bottom of the 1T1M crossbar, we need a multiplexer and a multiplexer decoder to select two columns in this example (in alternative implementations more columns could be selected to allow for batch updates);
4. The output currents from two columns are sent to a comparator to perform a threshold function.

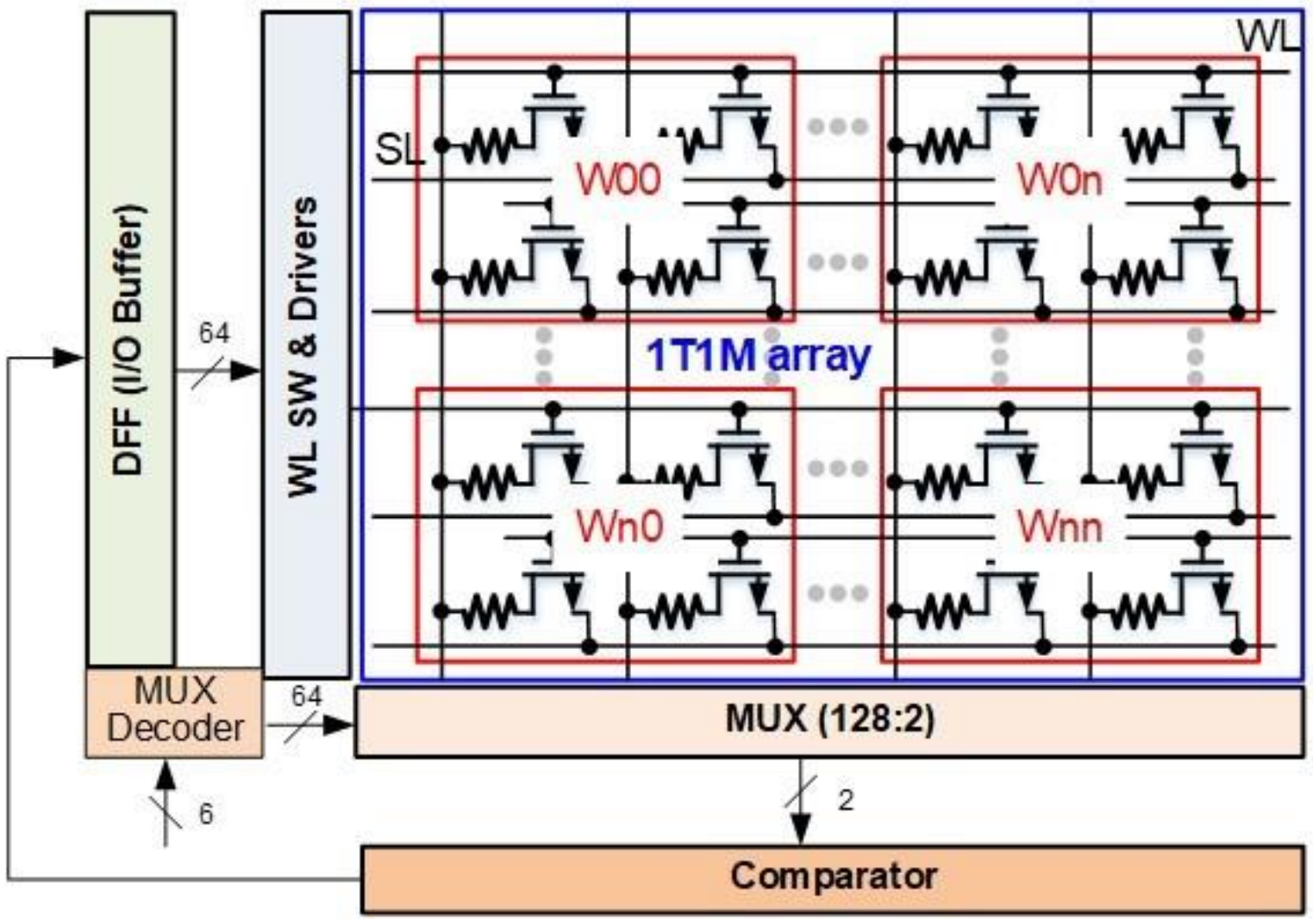

**Figure S4.** Circuit diagram of a 128×128 memristor crossbar for a Hopfield Neural Network.

All the area and energy values of the circuits in Figure S4 are summarized in Table S1. We use NeuroSim+[73] at 32 nm to model the area and energy for all the circuits required in Figure S4. The memristor cell size is assumed to be 12 $F^2$ (F = 32 nm). To estimate total energy or power values, we consider 100 cycles of operation, which may vary depending on the problem, accuracy of solution, etc. The 128×128 memristor crossbar is the dominant contribution to the power consumption. In the evaluation of this table, we assume that $R_{on}$ = 100 kΩ and $R_{off}$ = 1 MΩ, with the percentage of ON/OFF cells being 50% (in correspondence with the principle illustrated in Figure S3). Note that for a dense Max-Cut this is an overestimation of the power consumption by 2×, as 50% of the weights are zero there. We provide the power consumption for three different scenarios: having the whole matrix active, having a single column active, and having 10 columns active. The latter case corresponds to the batch update mechanism used in Figure 5 (which would only require a slight modification of the architecture in Figure S3).

In Table1, when mentioning the power consumption for dense unweighted Max-Cut for $N = 60$, we use the value of the energy consumption per clock cycle provided in Table S1 for the $N = 64$ system (resulting in 4 redundant rows and columns in the crossbar, some of which might be used for intrinsic noise generation), assuming a 1 GHz clock cycle and a 10 column batch update (the 21 $\mu$W leakage power is negligible). In addition, we include a 2× overhead, which includes a conservative estimate of the additional power needed to cool the memristor crossbar array. This 2× overhead estimate originates from the notion that the power consumption for the CPU solution, 20 W/core, is based on the total power consumption of a data center[42], divided by its number of cores. Equivalently, for a HPC-type of infrastructure consisting of mem-HNN cores, we would need to include the overhead included in the power usage effectiveness (PUE) of the HPC infrastructure. On average, typical PUE values have recently been decreasing[74] to ~1.6. In addition, the overhead due to the power consumption of cooling fans (not included in the PUE) needs to be accounted for as well, resulting in a total overhead of 2× . In contrast to the 20 W/core value for the CPUs, we have not explicitly incorporated the power consumption of the non-CPU compute infrastructure (i.e., network, storage, ...). Whereas this might result in a comparative overestimate of the power consumption of the CPU (<50%), this choice is justified given that the $N = 60$ benchmark task used in Table1is sufficiently small to run on a single accelerator-instance for all the technology entrees. Similar to what has been done for the AI-based matrix-vector product applications based on memristor-crossbars[25,40] , in future work, it can be addressed which multi-tile architecture would result in an optimal power consumption for the mem-HNN system for large-scale problems.

To verify the feasibility of the 1 GHz clock frequency, we performed simulations to determine the delay of the critical path during read or evaluation operation. Based on the delays of the different subcomponents of the proposed mem-HNN architecture, the critical path delay can be inferred to be 0.526 ns, which is well below 1 ns.

**Table S1.** Area and energy breakdowns for circuits in Figure S4. The total latency for one computing cycle of a 128 x 128 memristor crossbar array was found to be 128 ns.

| Component | **Area ($\mu m^2$)** | **Latency (ns)** | **Energy (pJ)** | **Leakage power ($\mu$W)** |
|---|---|---|---|---|
| **I/O buffer** | 245.12 | 1 | 0.023 | 1.5808 |
| **SW matrix & drivers** | 175.32 | 0.501 | 0.270 | 3.2467 |
| **Memristor crossbar (full)** | 201.30 | 0.01 | 198.10 | 0 |
| **Memristor crossbar (10 col.)** | 201.30 | 0.01 | 30.953 | 0 |
| **Memristor crossbar (1 col.)** | 201.30 | 0.01 | 3.095 | 0 |
| **MUX** | 939.52 | 0.0040 | 0.013 | 0 |
| **MUX Decoder (6-bit)** | 340.57 | 0.1237 | 0.110 | 16.37 |
| **Comparator** | 3.49 | 0.025 | 0.052 | 0.0062 |
| **Total (full-matrix)** | 1905.3225 | 1 | 228.021 | 21.2037 |
| **Total (10 col. active)** | 1905.3225 | 1 | 60.874 | 21.2037 |
| **Total (1 col. active)** | 1905.3225 | 1 | 33.016 | 21.2037 |

### 1.6 Matrix-vector multiplication errors due to crossbar noise

Figures S5a-b show more details of the simulation results used to support Figure 4c. We performed simulations of a vast ensemble of analog computations performed in memristor crossbar arrays of size 32×32 (Figure 4a) and 128×128 (Figure 4b). The analog computations in the 32×32 array give results very close to ideal results in such a way that there is no overlap of neighboring distributions. In other words, the error distribution (see inset figure) has a standard deviation that is sufficiently below 0.5 that any result can be rounded to the nearest integer and will match the exact integer result. In contrast, 128×128 arrays have large enough errors that simple rounding will not match the exact results, the error standard deviation (inset) can be much greater than 0.5.

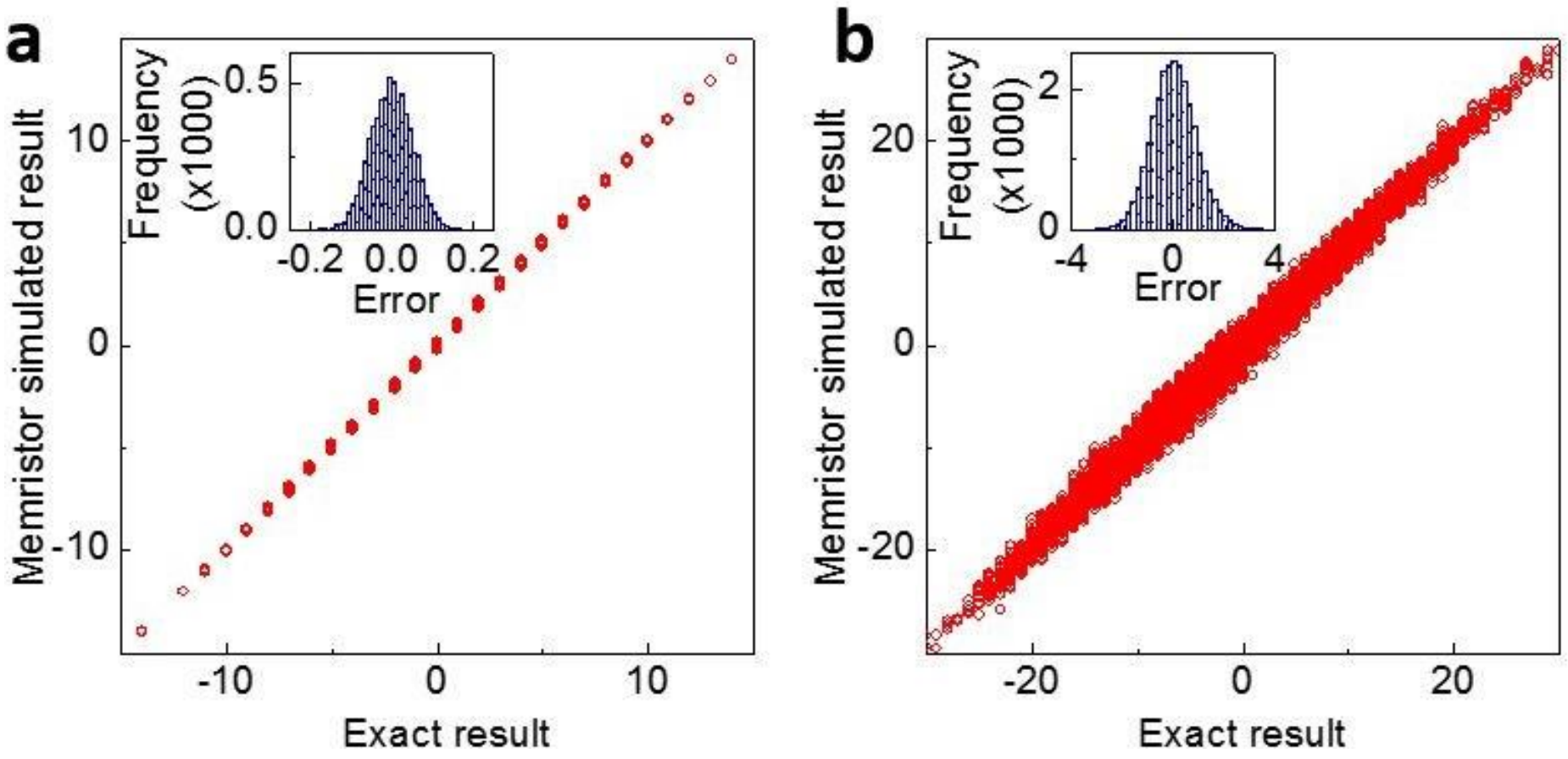

**Figure S5.** a) Computations in arrays of size 32×32 have non-overlapping analog results and error distributions (inset) lower than 0.5, matching digital results when rounded to the nearest integer. b) Larger arrays of 128×128 increase these error distributions such that analog output results can deviate more substantially from ideal integer results.

## 1.7 Additional details on the experimental results

The memristor crossbar array used to program the weights of the target problem was composed of memristors made using hafnium oxide as the active material, thereby operating via the oxygen-vacancy migration mechanism. Within the 128× 64 array of memristors, some rows and columns were identified to be open (not connected thereby no current flowing) or short (all devices within the row/column appear permanently in a very low resistance). We excluded the shorted rows and columns from the active parts of the array to be programmed. We included the open rows and columns by programming corresponding rows and columns within the array to a high resistance state, such that the programmed array was symmetric, a necessity of the HNN weights matrix. This process of altering the array to account for the non-ideal devices resulted in modification of Graph 0 in the Biq Mac Library to a problem that was slightly different from the one used in the simulations in the rest of the manuscript. The programmed weights and the corresponding distribution of the conductances are shown in Figure S6. Due to limitations of the experimentally available memristor crossbar chip, we did not yet implement the full architecture used in our simulations (shown in Figure S3), but we opted for an implementation in which one matrix element corresponds to a single memristor in the crossbar. Since the weights matrix for the max-cut problem contains 0's and -1's, the weights matrix programmed into the memristor crossbar consisted of either low ('0') or high ('1') conductances, and the resulting current from the vector-matrix multiplications were flipped in sign to account for the negative sign on the high conductance states. Additional details of fabrication, programming and operation of the array are given elsewhere.[37] Noise was introduced in the control software in a way identical to the algorithms in the simulations.

## 1.8 Additional context for the comparison of the different accelerators

This section provides additional information about Table 1 that could not be included in the main manuscript due to space constraints.

### 1.8.1 *Emerging technologies not included in Table 1*

Interestingly, recently, another promising digital annealing technology, has been proposed, namely an FPGA implementation of an optimized Boltzmann sampling algorithm called a Digital Annealer[9]. Impressively, the DA has been applied to Max-Cut with non-binary weights with a 16 bit resolution, even outperforming software algorithms that have a higher bit resolution[9]. We expect similar benefits for the mem-HNN system. Moreover, by embedding the crossbars inside the mem-HNN in an appropriate tiled architecture, the effective bit-resolution can be increased in future work[25]. Unfortunately, so far, the performance of the DA for the dense Max-Cut on $N = 60$ hasn't been reported and, as it behaves similar in power and speed to the CPU and GPU approach, it is not included in the benchmark table in the main manuscript. Similarly, the recent work on a CMOS annealer reported in[63,64] has not investigated its performance for the dense Max-Cut benchmark task and, as it only allows for sparse connectivity (hence bad scaling of the success probability and time-to-solution), we chose to not include it in the benchmark table due to place constraints.

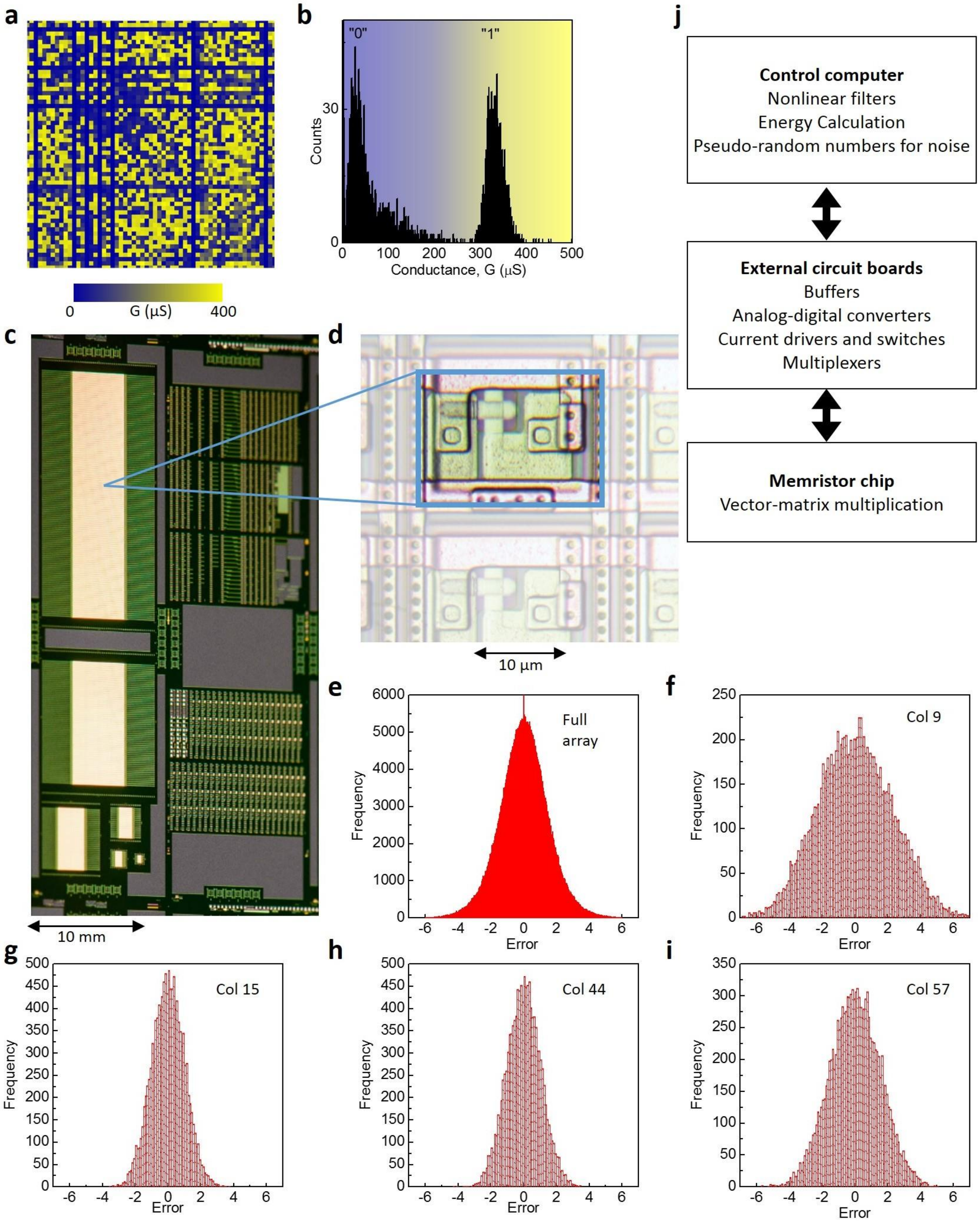

**Figure S6.** a) Spatial map of the conductances of the programmed memristor array. b) Histogram of the conductances from (a), displaying the distinct distributions corresponding to 0's and 1's in a). c) Magnified photograph of a typical memristor array chip, along with d) a further magnified picture of a single (1T1R) crossbar device. e) Histogram of errors from the experimental array. f)-i) Histograms of errors from four randomly selected columns from the experimental array. j) Schematic layout of the hardware functions in the implementation of the mem-HNN.

#### 1.8.2 Caveat about the power consumption numbers for quantum and CIM systems

In the case of D-wave's quantum annealing accelerator, 25 kW of power is needed to operate the system, of which the majority is used for cooling. This power envelope is expected to stay constant when the number of nodes scales into the tens of thousands, as the cooling is currently over-provisioned[75]. The idea is that, for some applications, a limited set of qubits might have a computational performance that cannot be obtained with a traditional high performance computing (HPC) system, even though HPC systems use 100-1000× more power. Due to this large fixed offset, when estimating the energy efficiency of quantum annealing based on the performance of current available small system sizes, the potential efficiency will be underestimated. Similarly, the current CIM implementations are just proof of concept implementations and are hence not yet optimized for energy efficiency.

#### 1.8.3 Benchmark simulations on a CPU using parallel tempering

Previous research has demonstrated that running parallel tempering (PT) on an Intel Xeon CPU E5-1650 v2 (3.50GHz) as a digital benchmark for the dense Max-Cut problem, reaches comparable time-to-solution values (TTS) as the fiber-based CIM proposed by Stanford and NTT[6]. Even though the power budget (thermal design power 130 W, actual estimated power 20 W) is lower than the power budget of the GPU (250 W), due to the stronger reduction in time-to-solution when using a GPU, running the noisy mean field annealing algorithm proposed by D-wave on a GPU[10] is outperforming the parallel tempering approach when run on a CPU.

The Max-Cut problem instances used in this work, are either taken from the Big-Mac Library $N = 60, 80, 100$, or randomly generated using the same problem definition as the dense Max-Cut problem proposed in the CIM analysis[6]. As an additional check on the difficulty of the specific Max-Cut instances used in Figure 5, we reran these very same problem instances on a CPU, and compared those with the previously published time-to-solution numbers for a CPU obtained[6] in the CPU-CIM comparison (Figure S4). This CPU reference performance is comparable with the performance of NTT's parallelized version of the fiber-based CIM.

In the experiment for this paper, an Intel(R) Core(TM) i7-3720QM CPU @ 2.60GHz was used. Similar to the CIM-paper, all the CPU results included here are provided by Salvatore Mandrà, using the same implementation of parallel tempering in the NASA/TAMU Unified Framework for Optimization (UFO). The algorithm (PT@UFO) has been executed on a single core. Despite the lower clock speed, the run-time analysis is still consistent with the previous analysis published in the NTT/Stanford paper.

In Figure S4, besides the reference data used in the Stanford/NTT paper, we include two different curves for the set of simulations ran on our own set of Max-Cut-problems: run-time without initialization time and run-time including the initialization time. Roughly speaking, the initialization time consists in the allocation of the arrays and the generation of all the random numbers that will be used in the simulation (all random numbers are cached and stored in memory). Hence, the run-time corresponds to the sole optimization of the Max-Cut problem. Typically, for larger systems, this initialization time is so fast that it is negligible compared to the run-time. However, for the rather small instances ($N < 100$) discussed in this paper, this is not the case. Importantly, random number generation happens intrinsicly in the proposed hardware for the mem-HNN system, the CIM system or the quantum system, which is one of the advantages of using analog hardware in accelerators. However, as the percentage of time spent initializing becomes negligible for larger systems, we report the TTS value *without* initialization for the CPU in Table 1. No description of initialization-time has been provided in the discussion of the GPU results[10].

A reasonable number for a single core energy consumption[42,44] is 20 W/core, this is the value used in Table 1. Additionally, the *TTS* for for the CPU benchmark can be improved by running independent processes in parallel, i.e., the *TTS* for PT@UFO can be reduced by the optimal number of repetitions $N_{rep} = TTS/T_{\mathrm{ann}}$ by running annealing instances on $N_{rep}$ CPU cores (for the small $N = 60$ task, the median $N_{rep} = 1$, so parallelization is not beneficial yet). Importantly, if the overhead of parallelization is negligible, this should have no significant effect on the energy to solution value listed in Table 1.

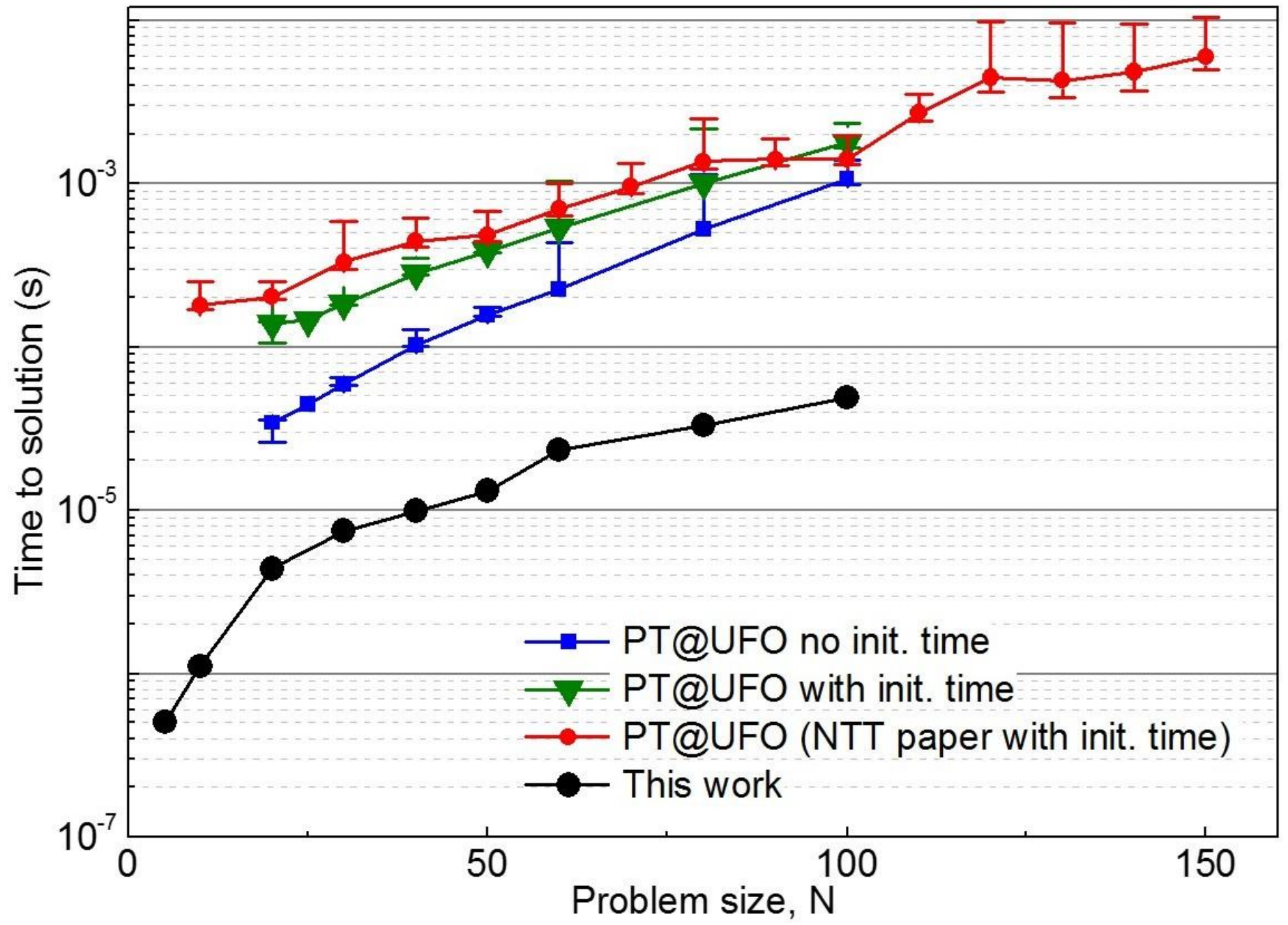

**Figure S7.** Time-to-solution (*TTS*) benchmark results for dense Max-Cut obtained by running parallel tempering on a CPU: (red circle marker) reference data obtained from the Stanford/NTT-paper[6], using the same problem definition, but different instances as in this paper; (triangle marker) CPU run-time on the same problem instances as used in the mem-HNN simulations, including initialization time; (square marker) CPU run-time on our problem set, without including the initialization time. Black circle markers are our results from Figure 5b, for 1000 cycles, using sequential runs (i.e., the worst case scenario shown in Figure 5b).

The time-to-solution in Table 1 for PT@UFO is calculated by running 1000 simulations of PT, with random seeds. For each simulation, the algorithm stops after the required number of sweeps to obtain the optimal solution, which we call the 'run-time'. For a given problem size $N$ and problem instance $i$, due to the different random seeds, every run-time is different and by calculating the cumulative distribution of all the obtained run-times, combined with a bootstrapping procedure, the optimal runtime $RT_{N,i}$ (= $T_{annealing}$) can be estimated that minimizes the time-to-solution $TTS_{N,i}$ for that task. The time-to-solution $TTS_N$ for a given size $N$ can then be obtained by using the same median value of $RT_N = \text{median}(RT_{N,i})$ for all the instances and by subsequently calculating for every task $i$ (with size $N$) the success probability $p_{sr,i}$ that a simulation for this task can be been completed in less time than $RT_N$ . Finally, these success probabilities $p_{sr,i}$ can be used to calculate for every size the median of the corresponding TTS-estimates (based on the number of repetitions $N_{rep,i} = \log(1 - p_{sr,i})/\log(1 - 0.99)$) for all these instances. Specifically, for $N$ = 60, $N_{rep} = 1$, consequently, as reflected in Table 1, $TTS_{60} = RT_{60} = T_{\text{ann}}$.

#### 1.8.4 Power consumption estimate for the GPU data

In Table 1, when including the energy efficiency of the GPU, we are using the 250 W as an upper bound for the thermal power consumption. Similar to the CPU scenario, where the thermal design power is 130 W, but in practice the power consumption during operation can be assumed to be[42,44] $\sim 20$ W/core, we assume the effective power of the GPU can be a factor 10× lower than the power budget estimate.

#### 1.8.5 Run-time analysis

Although the mem-HNN, the CIM and NMFA@GPU have a similar clock frequency (in the case of the CIM we pick the repetition rate of the pulsed laser as fundamental clock unit), and a similar time-to-solution scaling, there are still important differences in performance. Before we dive into the speed and energy efficiency specifics, it is worth noting the differences in modus operandi: the algorithm on the GPU can update the nodes synchronously, whereas the CIM system is purely asynchronous (one node gets updated per pump pulse), and, as explained in the discussion of Figure 5, the mem-HNN is somewhere in-between: we notice better performance in success probability if we update the nodes in batches $N_b = 10$ instead of node-by-node or completely synchronous. For the CIM system, doing updates in batches would be possible by using multiple phase-sensitive amplifiers in parallel, but that would be a drastic increase in system complexity. Given that traditional discrete-time Hopfield algorithms are known to have better performance when using asynchronous updates, the ability for the noisy mean-field algorithm to work with in essence synchronous updates can be surprising. However, the asynchronous updates were a mathematical trick to mimic noisy continuous-time Hopfield neural networks, where updates do occur ‘synchronously’. The noisy mean field algorithm can be seen as another way of modeling noise-induced annealing effects in a discrete-time algorithm that allows for synchronous updates. These choices do have important consequences on the eventual time-to-solution. One annealing run will consist of a certain number of cycles, and, for the mem-HNN, one cycle will take $\frac{N}{N_b} T_{clock}$. For the CIM, $N_b = 1$ and there is an additional overhead on the order of $1.6$ - $2.5$ related to additional pulses required for stabilizing the fiber feedback loop, increasing the annealing time. For the GPU the calculation time also dwells on the calculation of the matrix-vector products in the algorithm. In the DA implementation, only one node is updated per iteration, but due to its parallel-trial scheme the number of required cycles per run is reduced. Finally, despite the slightly higher clock cycle of the CPU, the parallel-tempering algorithm has a longer optimal annealing time than both the GPU-approach and the mem-HNN approach. This is probably related to its asynchronous update mechanism of the states of the Ising nodes.

As discussed in Sec. SM. 1.8.3, there is a procedure to obtain the optimal run-time/annealing time for the parallel tempering algorithm on the CPU. Similarly, in principle, the number of cycles per annealing run can be optimized for different problem sizes for the other classical technology implementations as well. However, the number of cycles used to obtain Figure S7 and Table 1 is fixed: 1000 for the GPU and CIM, 50 for mem-HNN (based on Figure 5b, resulting in a 20× difference). For a given annealing time, one run will then result in a success probability, and this success probability can be converted to a time-to-solution (i.e., the duration required to hit the optimal solution with 99% probability). Given the same clock frequency for GPU, CIM and mem-HNN, the shorter annealing time of the mem-HNN compared to the CIM can be solely explained by the stabilization pulse overhead for the CIM, the choice to use 20 less cycles per run for the mem-HNN, and the CIM’s current inability to update in batches. The GPU’s scaling of the annealing time versus problem size is less straight forward to explain, as it also scales with the (dense) matrix-vector multiplication, hence, for large problem sizes, we expect a quadratic dependence (if the available CUDA-cores is assumed to be fixed).

#### 1.8.6 Run-time versus success probability trade-off

The meta-parameters of the algorithm in the case of the noisy mean field annealing implementation, and the physics and operation regime of the hardware accelerators will then determine the success probability for a given annealing time. For instance, the success probability for the mem-HNN is lower than the success probability of the CIM, which results in the need to have more runs for the mem-HNN. However, due to the shorter annealing time, the absolute time-to-solution is still better. Obviously, a similar optimization routine based on the trade-off between allowing for a lower success probability to obtain an overall shorter time-to-solution could also be performed on the CIM system. Importantly, this wouldn’t make up for the overhead time required for stabilization of the feedback loop and the current lack of possibilities to use batches in that technology. Also, whereas the mem-HNN benchmark numbers are just based on (experimentally grounded) simulation, the CIM is an actual experimental implementation. In these experiments, there is also overhead for readout and postprocessing of data[6], which is currently not included in the table.

#### 1.8.7 Technology nodes for the electronic implementations

We assumed a 32 nm node for the simulations of the mem-HNN, which does not have access yet to Finfet transistors and is less advanced than the nodes for the CPU and GPU experiments, where a 22 nm and 16 nm node is used, respectively. Choosing a more advanced technology node for the mem-HNN architecture would improve its energy-efficiency even more.

#### 1.8.8 Outlook for optical and quantum technologies

Finally, we want to give an outlook on the future prospects of the involved technologies.

For the CIM, there is substantial room for improvement by either reducing the overhead required for stabilization of the feedback loop or increasing the effective clock frequency and/or by allowing for batch updates using several parallel phase-sensitive amplifiers. One other promising alternative is to integrate the coherent Ising machine into a photonic integrated circuit[7,8]. This would result in relatively lower cost, potential for scalability, lower energy consumption and higher clock speeds.

As for potential improvements in quantum annealing, demonstrating larger node sizes will result in an increased energy efficiency, but without changes in the interconnectivity, we do not expect this technology to become competitive with the other listed approaches for the currently studied benchmark task. Consequently, as mentioned in Ref.[6], there are also efforts on allowing for all-to-all connectivity in quantum annealing implementations, which would result in similar scaling as the other classical approaches, but this technology development is non-trivial[62,76]. Besides increasing the number of qubits in a single D-wave system, to drastically improve the energy-efficiency of the quantum annealer far more efficient cryogenic cooling techniques need to be developed, but it is currently unclear which techniques might be suitable for this task[42].